\documentclass[sigconf]{acmart} 
\usepackage{wrapfig}
\usepackage{caption}
\usepackage{subcaption}
\usepackage[utf8]{inputenc} 
\usepackage[normalem]{ulem}
\usepackage{booktabs}
\usepackage{enumitem}

\AtBeginDocument{%
  \providecommand\BibTeX{{%
    \normalfont B\kern-0.5em{\scshape i\kern-0.25em b}\kern-0.8em\TeX}}}

\copyrightyear{2021}
\acmYear{2021}
\setcopyright{acmlicensed}\acmConference[CHI '21 Extended Abstracts]{CHI Conference on Human Factors in Computing Systems Extended Abstracts}{May 8--13, 2021}{Yokohama, Japan}
\acmBooktitle{CHI Conference on Human Factors in Computing Systems Extended Abstracts (CHI '21 Extended Abstracts), May 8--13, 2021, Yokohama, Japan}
\acmDOI{10.1145/3411763.3451768}
\acmISBN{978-1-4503-8095-9/21/05}

\begin{document}

\title[A Platform for Open Global Air Quality Monitoring]{Towards an Open Global Air Quality Monitoring Platform to Assess Children's Exposure to Air Pollutants in the Light of COVID-19 Lockdowns}

%


\author{Christina Last}
\email{christina@skwire.co.uk}
\authornote{Equal Contribution}
\affiliation{%
  \institution{Skwire}
  \country{UK}
}

\author{Prithviraj Pramanik}
\email{prithvirajpramanik@yahoo.co.in}
\authornotemark[1]
\authornote{Corresponding Author}
\orcid{0000-0002-5656-6340}
\affiliation{%
  \institution{NIT Durgapur, India \& Discovery Partners Institute}
  \country{USA}
}

\author{Nikita Saini}
\email{nikitasain61@gmail.com}
\authornotemark[1]
\affiliation{%
  \institution{CitiGroup}
  \country{India}
}

\author{Akash Smaran Majety}
\email{akashsmaran@gmail.com}
\authornotemark[1]
\affiliation{%
  \institution{Humaine}
  \country{USA}
}

\author{Do-Hyung Kim}
\email{dokim@unicef.org}
\author{Manuel García-Herranz}
\email{mgarciaherranz@unicef.org}
\affiliation{%
  \institution{Office of Innovation, UNICEF}
  \country{USA}
}

\author{Subhabrata Majumdar}
\email{subho@att.com}
\affiliation{%
  \institution{AT\&T Inc.}
  \country{USA}
}

\renewcommand{\shortauthors}{}

\begin{abstract}
  This ongoing work attempts to understand and address the requirements of UNICEF, a leading organization working in children's welfare, where they aim to tackle the problem of air quality for children at a global level. We are motivated by the lack of a proper model to account for heavily fluctuating air quality levels across the world in the wake of the COVID-19 pandemic, leading to uncertainty among public health professionals on the exact levels of children's exposure to air pollutants. We create an initial model as per the agency's requirement to generate insights through a combination of virtual meetups and online presentations. Our research team comprised of UNICEF's researchers and a group of volunteer data scientists. The presentations were delivered to a number of scientists and domain experts from UNICEF and community champions working with open data. We highlight their feedback and possible avenues to develop this research further.  
\end{abstract}




\maketitle

\section{Introduction}\label{sec:intro}
There is a strong link between human health and exposure to high levels of air pollution \cite{Kampa,Kim}. Deteriorating air quality has thus been a growing concern over the last decade \cite{who2016, brunekreef2002air}, in particular within urban agglomerations \cite{cohen2004urban,schwela2000air}. In order to monitor the temporal and spatial variability in air pollution across regions where ground sensors are sparse, there is a need to develop data-driven approaches to predicting air pollution levels. Several such prediction methods use historic and current data from existing air quality monitoring stations \cite{Uair}. Despite this progress, there is a need to advance current modelling approaches to account for the variation introduced from COVID-19 lockdown events at a global scale, with particular focus on modelling air pollution concentration in regions with sparse coverage of ground-based air quality monitoring sensors.

Long-term exposure to fine particulate matter (especially, PM2.5, i.e. particles with a diameter less than 2.5$\mu$m) are estimated to cause $\sim$8 million excess deaths annually. The impact of air pollution is felt more acutely by the young, with 1 in every 4 deaths under 5 years is directly or indirectly related to environmental risks \cite{who}. Globally, 93\% of children live in places where air pollution levels exceed World Health Organization (WHO) guidelines \cite{WHO_AQ}, defined as PM2.5 values above 10 $\mu g/m^3$  annual mean and 25 $\mu g/m^3$ 24-hour mean \cite{Venter}. Therefore, there is an increasing need to develop a model that explains temporal and spatial variability in air pollution concentrations to accurately assess the global children populations' exposure to high levels of air pollution. 

\begin{figure*}[t]
    \centering
    \begin{tabular}{cc}
    \includegraphics[width=.99\linewidth]{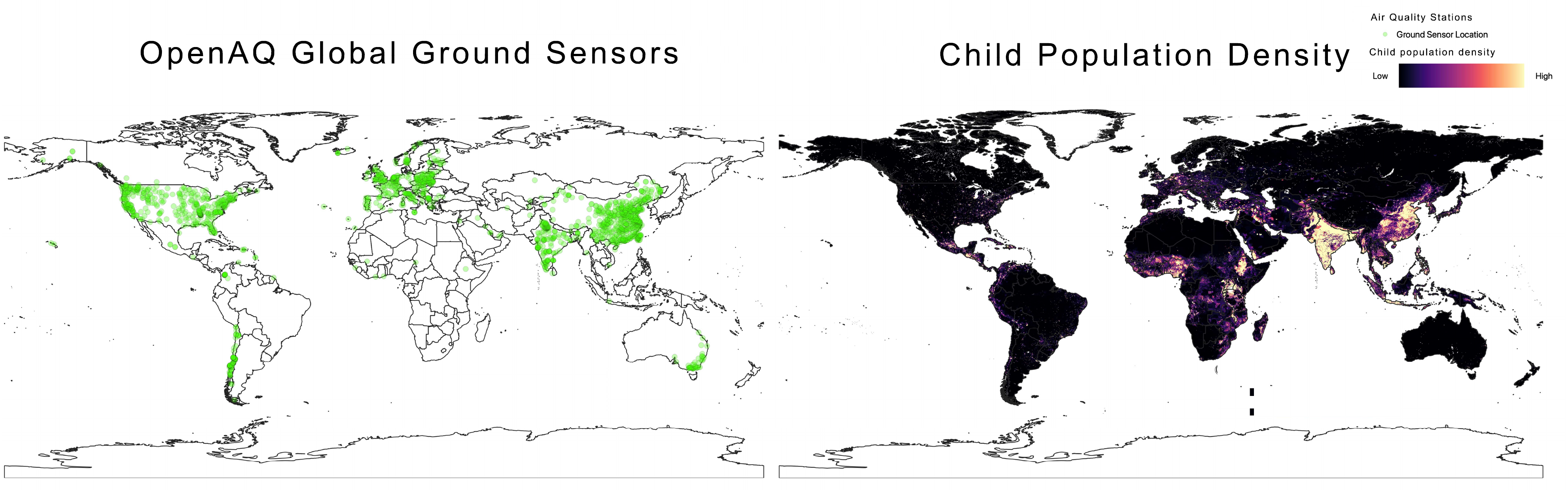} &\\ 
    \begin{minipage}{0.5\textwidth}
    \subcaption{\label{left}}
    \end{minipage}
    \begin{minipage}{0.5\textwidth}
    \subcaption{\label{right}}
    \end{minipage}
    \end{tabular}
    \caption{Global distribution of (a) Ground Level Air Quality Sensors, and (b) Child Population, Data source: \citet{silent}.}
    \label{fig:mapscompare}
\end{figure*}

\subsection{Motivation}
\label{subsec:motiv}
As seen in Figure~\ref{fig:mapscompare}a, ground-based air quality sensors are distributed worldwide heterogeneously, with particularly low concentration of sensors in West and East Africa. A comparison with the distribution of child population across the world (Figure \ref{fig:mapscompare}b) indicates that it is difficult to assess can children's exposure to high levels of air pollution in areas with scarce ground-based observations, such as in highly populous locations like Western Africa and the Great Rift Valley. Responding to this challenge, UNICEF\footnote{\url{https://www.unicef.org/}} and Solve for Good\footnote{\url{https://www.solveforgood.org/}} ---a non-profit foundation connecting philanthropic organizations with data science expertise---aim to to build an open global model for air quality prediction.

Children are likely to have more acute respiratory illnesses than adults when exposed to high levels of air pollution, often with long-term consequences \cite{Zvobgo,ALA}. Moreover, the detrimental effects of COVID-19 are  compounded by poor air quality. COVID-19 is a lung disorder, to which air pollutants have been a major contributor \cite{coker2020effects, isphording2020pandemic}. Motivated by such concerns---which are in tune with UNICEF's
focus on child health---the objectives of this collaborative research project are to:
\begin{enumerate}[leftmargin=*]  
    \item Build a global model that augments accurate but limited air pollution data collected using ground sensors with machine learning (ML) predictions using global-level but coarse satellite data as input features.
    \item Develop regional, "fine tuned" versions of the model to address inaccuracies within a global model to understand the variability of air pollution patterns across regions, specially with sparse ground sensors.
    \item Measure children’s exposure to high levels of air pollution, in particular where children are exposed to pollution concentrations exceeding the WHO pollution standards during  pre- and post-COVID-19 lockdown events.
    \item Educate UNICEF's regional offices, partner organisations and citizens on the benefits of data-driven air pollution modelling and use their feedback to improve our system.
\end{enumerate}

\begin{figure*}[t]
    \centering
    \includegraphics[width=0.7\linewidth]{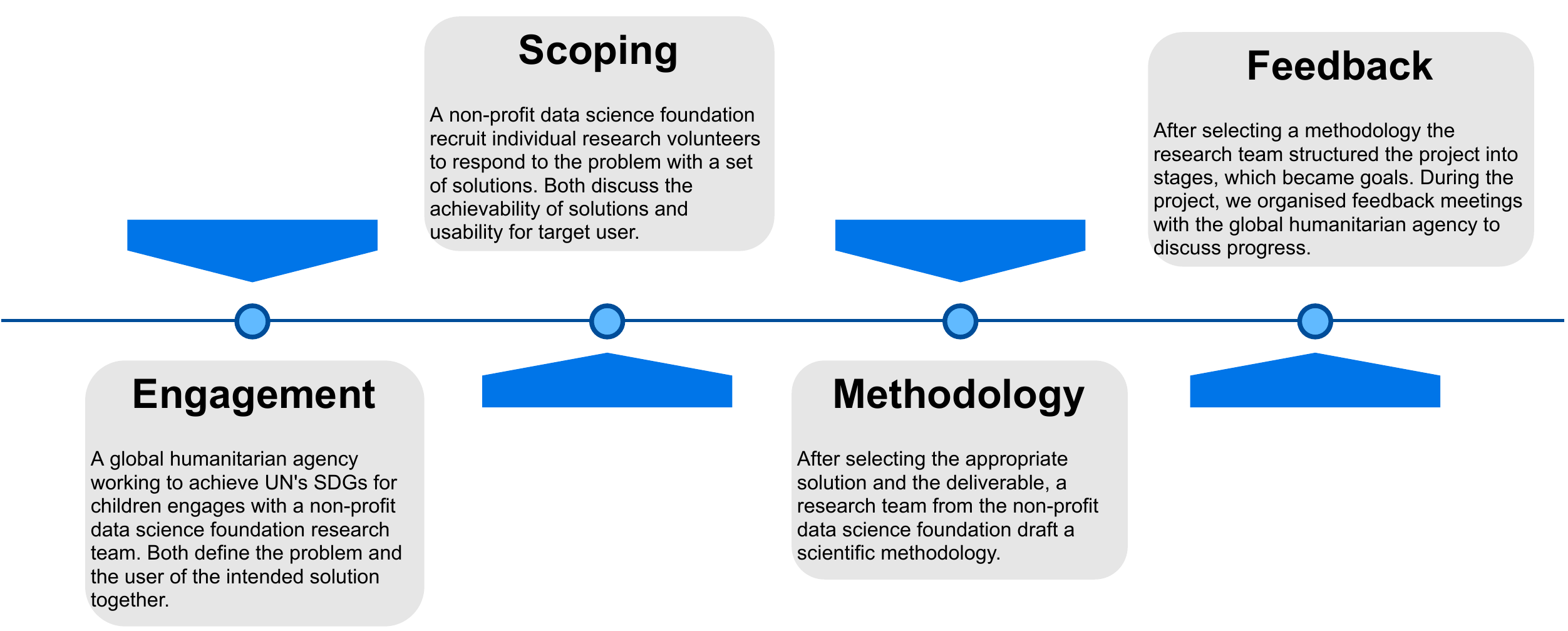}
    \caption{The project process 1. Identifying the problem, 2. Scoping a potential solution, 3. Developing a scientific methodology, and 4. Receiving feedback from project partner (UNICEF) and update project process.
}   \label{fig:Process_diagram}
\end{figure*}

\subsection{Project Workflow} 
In the first stages of the work, UNICEF approached Solve for Good with  challenge---that it is difficult for the UNICEF's regional offices, and local partner organisations to monitor air pollution at a local level, resulting in insufficient evidence to justify investing in air quality improvement projects. (Figure \ref{fig:Process_diagram}: Stage 1 - Engagement).

After discussing different aspects of this challenge, the project team scoped a solution to fit the time and resources available (Figure \ref{fig:Process_diagram}: Stage 2 - Scoping). The project scope focused selecting a research question that responded to the challenge, and leveraged the data science expertise in the Solve for Good team of volunteers. 

Following the scoping stage, UNICEF worked in collaboration with the Solve for Good researchers to develop a scientific methodology outlining the stages of data collection, pre-processing, analysis and visualisation (Figure \ref{fig:Process_diagram}: Stage 3 - Methodology). Key datasets were identified in this stage, such as satellite data and ground-based measurements of air quality (Section~\ref{sec:data}), and data such as the global child population and school locations.

During the development of the methodology, the project team integrated weekly feedback sessions to report on the project analysis (Figure \ref{fig:Process_diagram} Stage 4: Feedback), giving UNICEF officials the opportunity to inform the research efforts with feedback from regional partners, as well as tailoring to the research needs of UNICEF right from the beginning.

\section{Data}\label{sec:data}

In the interest of using open data, we used the data provided by OpenAQ\footnote{\url{https://openaq.org}}, and collected global air quality data from air quality monitoring stations across the globe (sample size $n$=1601), measuring ground-level PM2.5 concentrations from January 2019 to September 2020. These ground-level values would act as our output feature (target variable), which we predicted using a machine learning (ML) model trained on weekly averaged value for AOD, NO$_2$, Population Density and Precipitation (acquired from the Google Earth Engine\footnote{\url{https://earthengine.google.com}}) as input features (predictor variables). We collected the input features for the time frame 1st January 2019 to 11th September 2020 from the satellites given in Table \ref{tab:datatable}.

%


\begin{table*}[!htbp]
\centering
\scalebox{.7}{
\begin{tabular}{|c|l|l|l|l|}
\hline
\textbf{Data Name} &
  \multicolumn{1}{c|}{\textbf{Satellite Name}} &
  \multicolumn{1}{c|}{\textbf{Spatial Resolution}} &
  \multicolumn{1}{c|}{\textbf{\begin{tabular}[c]{@{}c@{}}Temporal \\ Resolution\end{tabular}}} &
  \multicolumn{1}{c|}{\textbf{Description}} \\ \hline
\textbf{Precipitation} &
  \begin{tabular}[c]{@{}l@{}}GSMaP Operational: Global Satellite Mapping \\ of Precipitation (Precipitation)\end{tabular} &
  0.1 Arc Degrees &
  Daily &
  \begin{tabular}[c]{@{}l@{}}Snapshot of hourly precipitation rate, \\ source: GSMaP Operational\end{tabular} \\ \hline
\textbf{AOD} &
  \begin{tabular}[c]{@{}l@{}}MCD19A2.006: Terra \& Aqua MAIAC Land \\ Aerosol Optical Depth Daily 1 Km (AOD)\end{tabular} &
  1000 Meters &
  Daily &
  \begin{tabular}[c]{@{}l@{}}Blue band (0.47 $\mu$m) aerosol optical \\ depth over land, source: MCD19A2.006\end{tabular} \\ \hline
\textbf{NO$_2$} &
  \begin{tabular}[c]{@{}l@{}}Sentinel-5P NRTI NO$_2$: Near Real-Time \\ Nitrogen Dioxide (NO$_2$)\end{tabular} &
  0.01 Arc Degrees &
  Daily &
  \begin{tabular}[c]{@{}l@{}}Total vertical column of NO$_2$ = ratio \\ of NO$_2$ slant column density and total air \\ mass factor, source: Sentinel-5P NRTI NO$_2$\end{tabular} \\ \hline
\textbf{Population Density} &
  \begin{tabular}[c]{@{}l@{}}GPWv411: Population Density Gridded \\ Population of the World (Population Density)\end{tabular} &
  30 Arc Seconds &
  Daily &
  \begin{tabular}[c]{@{}l@{}}The estimated number of persons per square\\  kilometer, source: GPWv411\end{tabular} \\ \hline
\end{tabular}}
    \caption{Different satellite data sources, along with brief descriptions.}
    \label{tab:datatable}
\end{table*}

\begin{figure*}[!htbp]
    \centering
    \includegraphics[width=.6\linewidth]{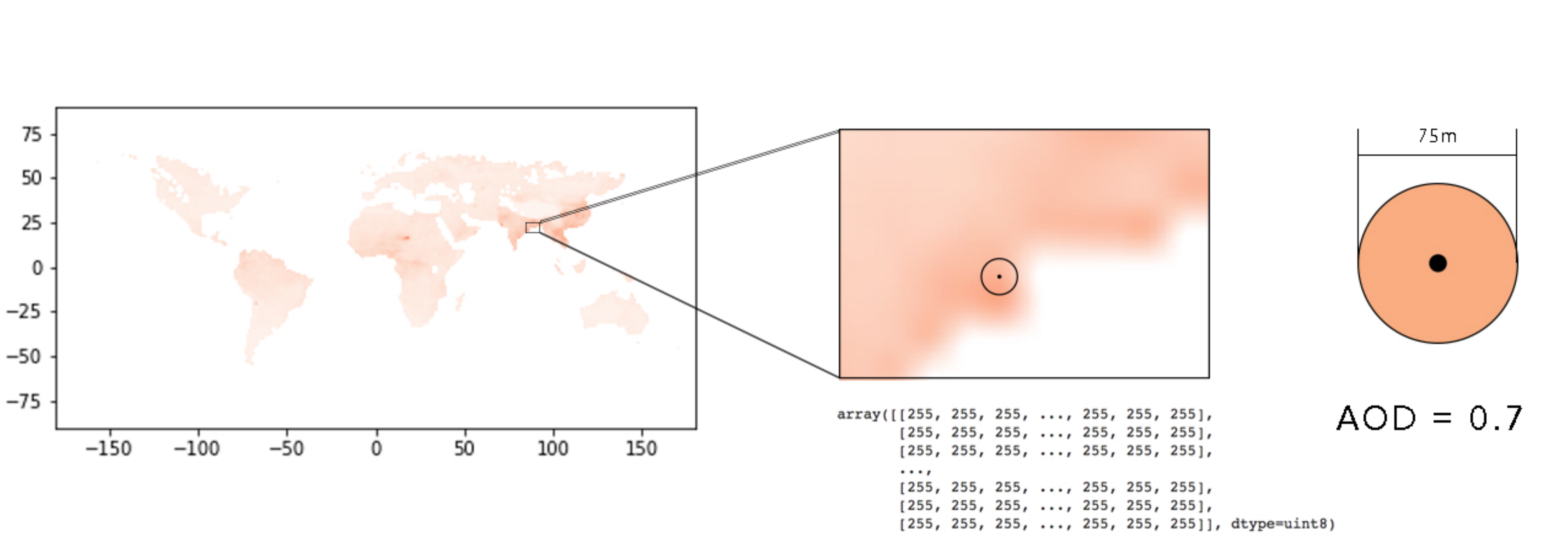}
    \includegraphics[width=.6\linewidth]{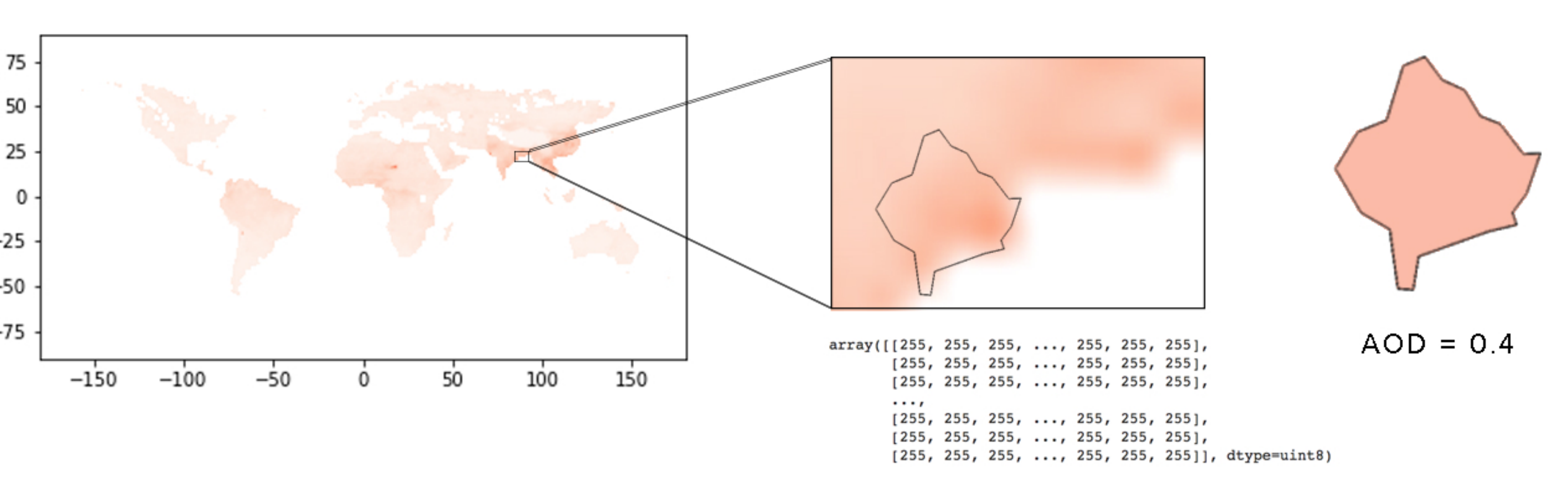}\\
    \caption{Top panel illustrates the "local" extraction of AOD from a 75m buffer around a point. Bottom panel illustrates the "city-level" extraction of AOD for each city globally. AOD = Aerosol Optical Depth is a well-known proxy for PM2.5 \cite{Kumar}.}
    \label{fig:2step}
\end{figure*}

\begin{figure*}[!htbp]
    \centering
    \scalebox{.8}{
    \begin{tabular}{cc}
    \includegraphics[width=.40\linewidth]{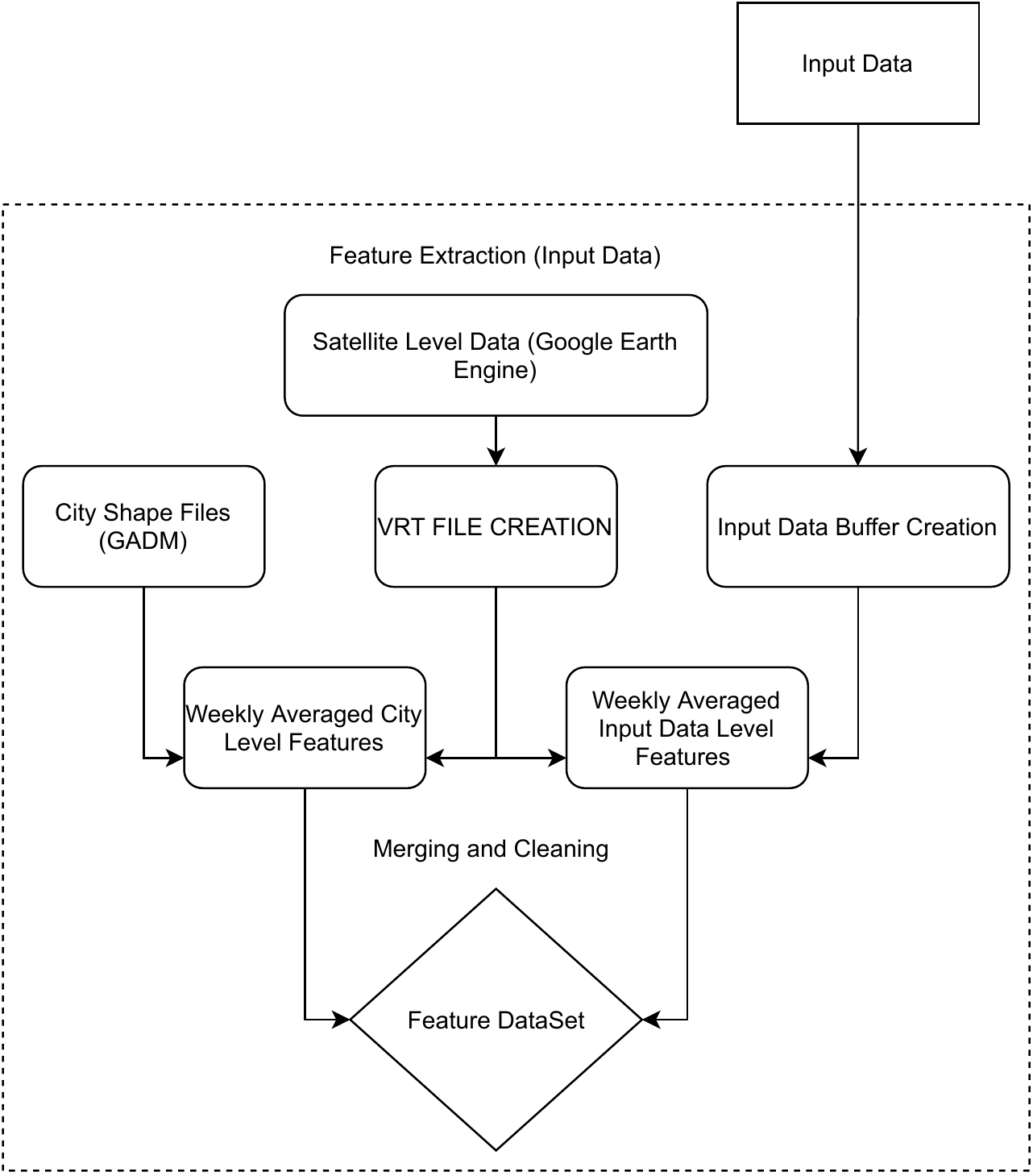} &
    \includegraphics[width=.25\linewidth]{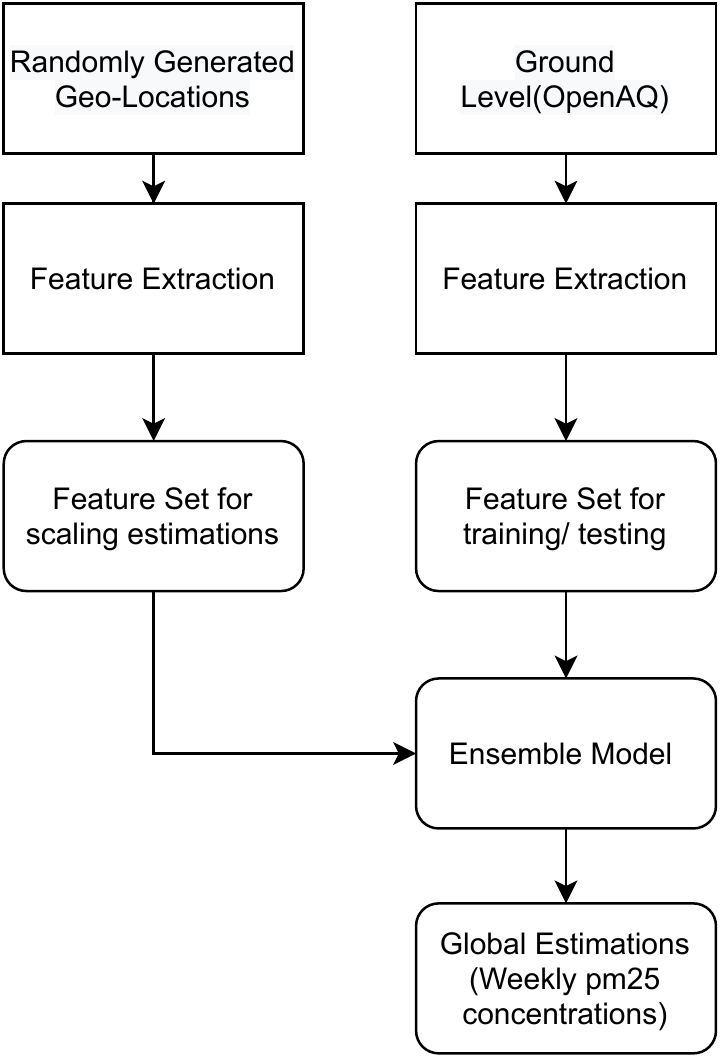}\\
    (a) & (b)
    \end{tabular}}
    \caption{Flow diagrams for (a) Data Preparation, and (b)  Modelling Process.}
    \label{fig:data_and_model}
\end{figure*}

Since the raw satellite data were collected at different temporal and spatial granularities (Table~\ref{tab:datatable}), we first standardized the data temporally and spatially to obtain our input features.  The PM2.5 values from the OpenAQ data had a mean (M) of 22.21 $\mu g/m^3$ with a standard deviation (SD) of 29.85. The data distribution had half of the values below 12.66 $\mu g/m^3$ and three fourths of the values under 27.74 $\mu g/m^3$. For output feature values, only PM2.5 values between 0 and 3000 were considered, and resampled on a weekly basis \footnote{Considering Monday as the starting day of the week\\} to calculate the weekly average PM2.5 concentration.  

Along with the satellite and PM2.5 measurement we considered the city wide lock-down information (Footnote: \ref{foot:Stringency}) from OxCGRT/ACAPS Lockdown Data (Footnote: \ref{foot:OxCGRT}, \ref{foot:ACAPS}) for 192 countries, but removed them later from the final feature set because of location heterogeneity issues (Further discussion in Section~\ref{sub:aux}).

We generated a VRT driver\footnote{A format driver for GDAL that allows a virtual GDAL dataset to be composed from the other GDAL datasets, \url{https://gdal.org/drivers/raster/vrt.html}} 
to enable faster processing of the vast amount of spatial data. We obtained weekly averages of the features at two different geographic levels---local and city-wide. At the local level, we derived the values using a local mask defined as a 75m buffer around each ground sensor location to capture intersecting pixel edges, and take their average. At the city level, we extracted the city-wide weekly averages using corresponding city masks. The final pre-processing step was to aggregate the sensor point data (along with latitude and longitude upto 4 decimal points) with the city and country identifiers. In Figure~\ref{fig:data_and_model}a, we summarize the full data preparation process.



\section{Methods}\label{sec:method}
\subsection{Model}
To achieve the first objective (Section~\ref{subsec:motiv}), we developed a stacked ensemble Machine Learning model (Figure~\ref{fig:data_and_model}b). The model enabled us to leverage the capabilities of a range of well-performing models to improve the accuracy of PM2.5 predictions over any individual model. We used LightGBM, XGBoost and Random Forest as the base models to generate predictions of weekly averaged PM2.5 concentrations. Base models were trained on 80\% of the OpenAQ PM2.5 concentration data. Predictions from base models were aggregated to be used as the input features for the meta-model. We used linear regression with a 5-fold cross-validation to train the meta-model.

To reduce the generalization errors of the individual learners, we used a stacking regressor (stacked base and meta models). Feature importances of the base learners showed high variability, with AOD being the most important feature in XGBoost and Random Forest, LightGBM on the other hand had average population density as the most important feature. An ensemble approach reduced this variance making the model more robust and stable in performance.

To validate the model at global level, we generated $10^6$ land-based locations \cite{Karin}, randomly distributed to ensure coverage of areas with little ground-based observation data of PM2.5 concentrations. We extracted satellite data using a VRT driver to generate features for these locations to predict the weekly average PM2.5 concentrations using our pre-trained ensemble model. Overall, our model achieved a median absolute error of  5.60 $\mu g/m^3$, with Root Mean Squared Error (RMSE) of 22.8 and $R^2 = 0.49$. On a more granular level, the model performed best in Myanmar (RMSE $= 9.9, R^2 = 0.8$) and worst in Ghana, attributed to the lack of training data. The model performed fairly well at locations in Guatemala, Iraq, Kyrgyzstan and Chile, with $R^2 >0.45$, and RMSE values in the range of 6-14.

\begin{figure*}[!htbp]
    \centering
    \scalebox{.8}{
    \begin{tabular}{cc}
    \includegraphics[width=0.5\textwidth]{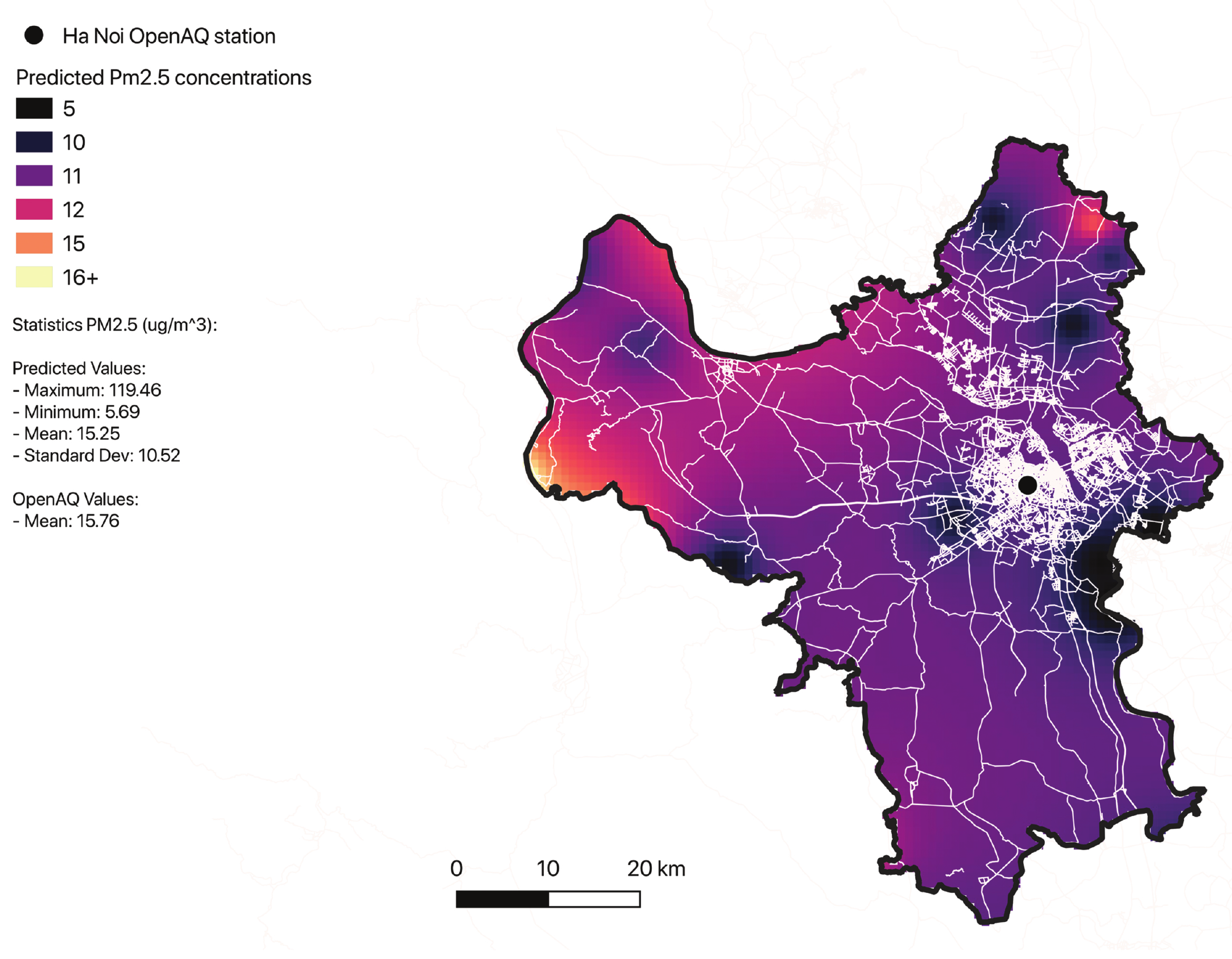} &
    
    \includegraphics[width=0.4\textwidth]{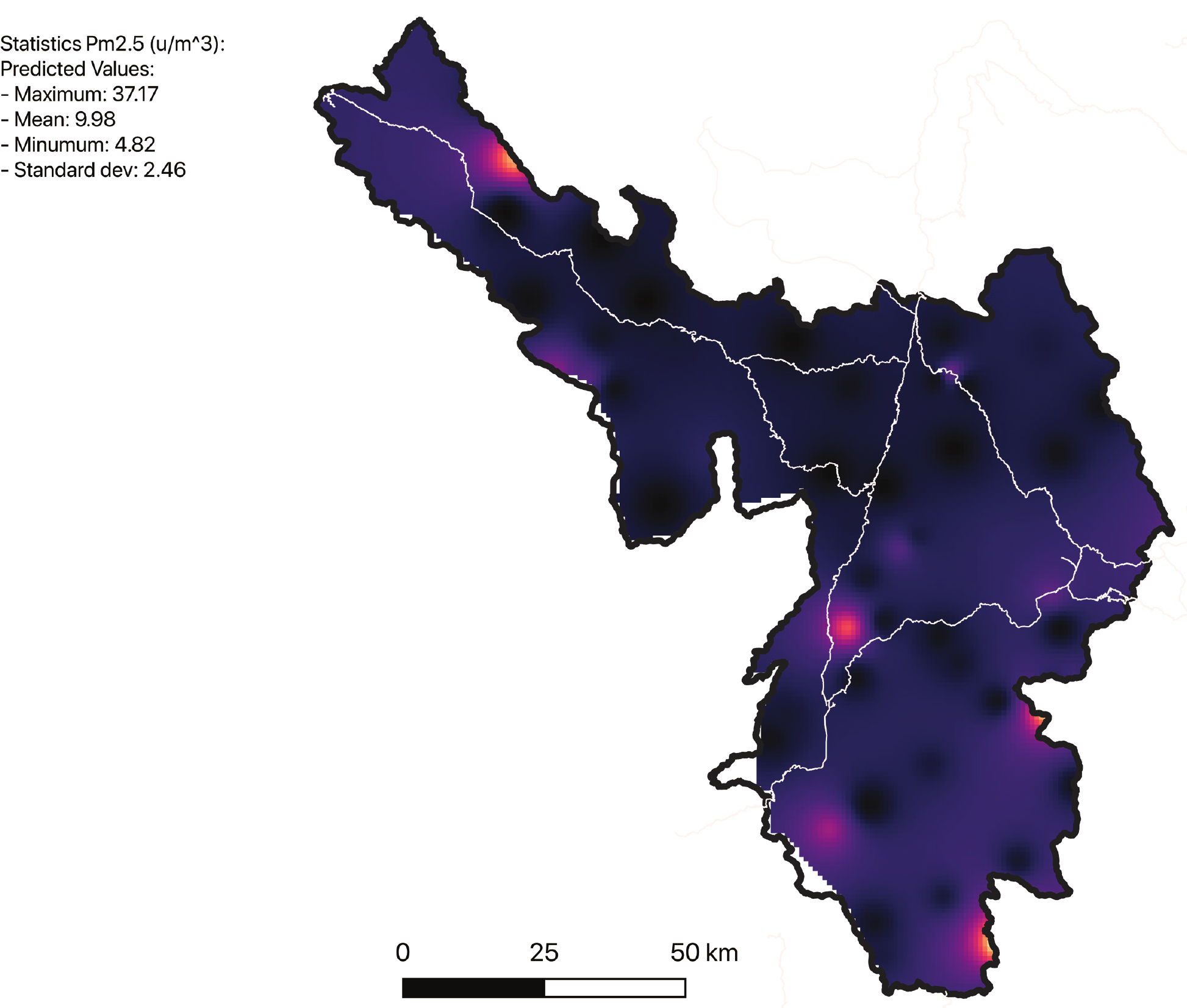}\\
    \centering
    (a) & (b)
    \end{tabular}}
    \caption{Comparative analysis of (a) H\`{a} N\^{o}i (an urbanised location in northern Vietnam) and (b) \DJ i\^{e}n Bi\^{e}n (a rural region in northern Vietnam) and their predicted PM2.5 concentrations}
    \label{fig:vietnam}
\end{figure*}

As an example, Figure \ref{fig:vietnam} describes the spatial variability of PM2.5 predictions the week beginning 20\textsuperscript{th} March 2020 in two locations in Vietnam. Figure \ref{fig:vietnam} (a) air pollution in H\`{a} N\^{o}i, with weekly average PM2.5 concentrations (M = 15.25 $\mu g/m^3$, SD = $\pm$10.52) in urban setting generates relatively high air pollution levels in comparison to the region of \DJ i\^{e}n Bi\^{e}n \ref{fig:vietnam} (b) (M = 9.95 $\mu g/m^3$ SD = $\pm$2.49) where the concentrations are much lower.

\subsection{Stakeholder involvement}

\begin{wrapfigure}[32]{l}{0.18\textwidth}

\centering
\includegraphics[width=.18\textwidth]{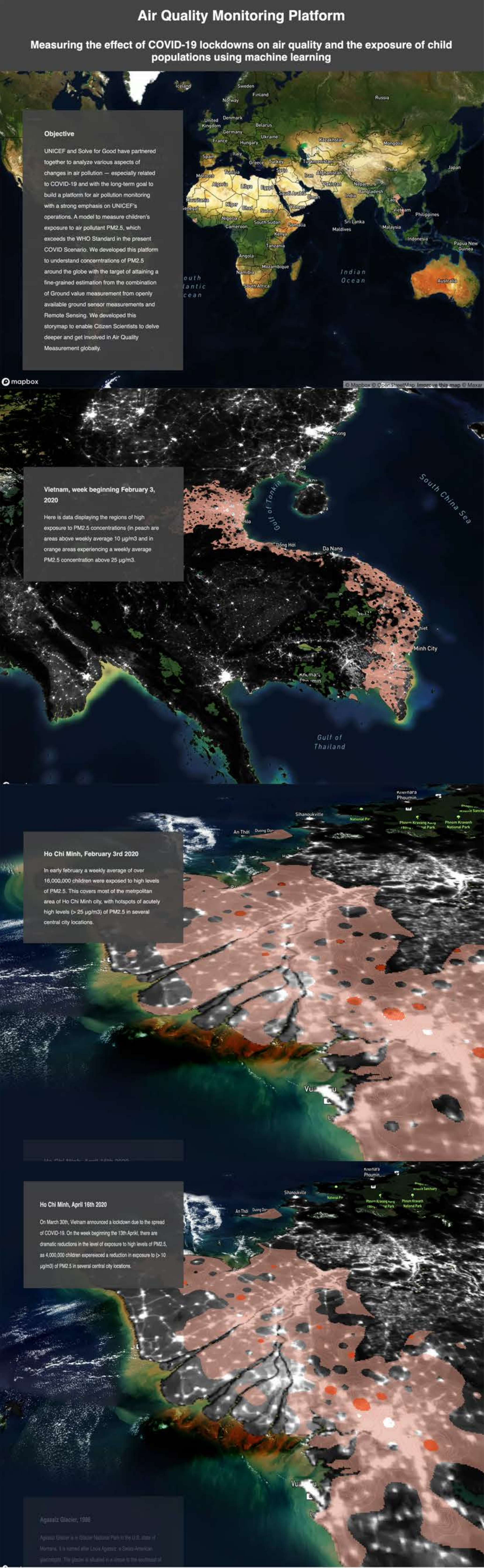}
\caption{Mapbox story map showing the results of the analysis.}
\label{fig:maps}
\end{wrapfigure}

In the first phase of the project process (Engagement, Figure~\ref{fig:Process_diagram}), our team consulted with 4 members from UNICEF
---1 climate and energy specialist, and 3 data scientists---to sketch out UNICEF's requirements and a timeline for the project. Two of the partner data scientists were involved in the second and third stages (Scoring and Methodology, respectively, in Figure~\ref{fig:Process_diagram}). UNICEF's ongoing involvement served two purposes: first, their domain knowledge helped data gathering, processing, and model building. Secondly, their continuous feedback during the weekly status meetings ensured that the final deliverable (global model and its outputs) was tailored to UNICEF's needs.

In the expert feedback and Q\&A session, we presented the methodology and results to 21 scientists and domain experts from UNICEF, and one open data and public policy expert from Solve for Good. To communicate the results of the research to UNICEF's regional offices and citizens (satisfying objective 4), we developed a story map (Figure~\ref{fig:maps}) using Mapbox, demonstrating the changes in predicted PM2.5 concentrations before and after COVID-19 lockdown events globally. The story map contains layers showing the child population density, and overlays polygons indicating geographic regions with PM2.5 concentrations above the WHO recommended limits (PM2.5 exposure of 10 $\mu$g/m\textsuperscript{3} annual mean or 25 $\mu$g/m\textsuperscript{3} 24-hour mean). Figure~\ref{fig:maps} indicates the reduction in exposure to high PM2.5 concentrations by comparing a pre-lockdown weekly average in early February 2020 to a week in mid-April 2020 for in Ho Chi Minh City, Vietnam. When presenting this story map in the Q\&A session, we used a between-subject study design to collate themes that emerged from the feedback received on implementation of the results. 

\section{Expert Feedback}
\label{sec:res}

A number of common themes emerged from the ensuing discussion and Q\&A on how to help translate the insights from our model into practical local-level policies.

\paragraph{Data scarcity: }
The model results indicate that scarce air quality monitoring stations (Figure~\ref{fig:mapscompare}a) reduce model accuracy, as predictions are less reliable at locations with no proximate air quality monitoring stations, or areas with high spatial or temporal variations of pollutant levels---often due to seasonal local factors such as festivals \cite{diwali} or agricultural practices \cite{cropburning}. In the absence of air quality monitoring stations in close proximity, ground or near ground-level air quality data can be obtained by collaborating with local corporations or nonprofits. For example, in line with objective 2 (to develop a regional, "fine tuned" versions of the model). In response to this challenge, a member of the expert feedback group suggested UNICEF's global field presence can leveraged to obtain more granular non-regulatory grade sensor, and  Plume Labs\footnote{\url{https://plumelabs.com}} agreed to share high-quality sensor data on PM2.5 concentrations across Lima, Peru.



        
\paragraph{Links to COVID-19 infection rates: }
Through a discussion with UNICEF's health experts, one significant suggestion to account for the association between COVID-19 incidences and poor air quality was to systematically correlate and compare administrative data from public health systems with levels of air quality in a geographic area pre- and post-COVID-19. This would help distinguish child cases of air pollution related respiratory conditions from COVID-19 cases. As we work towards a broader understanding of child respiratory health metrics and their connection to air quality in the wake of COVID-19, obtaining region-wise risk classifications based on observational public health data can help us translate local air quality predictions into concrete policy interventions.
\paragraph{Local heterogeneity of data and transparent predictions: }
Explainable methods should drive transparency in models that use features from local data and predictions from a global model such as ours (demonstration video \footnote{\url{https://youtu.be/2JVDRpLRX2k}}). Based on user needs, such explanations could use simple parametric models (such as linear regression), or techniques with post-hoc explanations \cite{Murdoch}.

\paragraph{Technical feedback: }
A prominent point brought forward by a open data and public policy expert from the non-profit foundation was the difficulty in validating PM2.5 measurement in regions which are completely missing from the training set.  In response to this feedback, we aim to create training and validation datasets to mimic local air quality patterns in regions that have sparse observational data. To this end, encouraging the participation of more regional offices could provide the technical team with local data sources. Additionally, instead of splitting the data randomly by location to obtain train/validation sets (current approach), a stratified sampling of locations can be undertaken---with sampling weights for a location as a function of the number of air quality monitoring stations within a certain radius. This can potentially diversify the test regions under consideration and improve model prediction quality in data-poor regions. Mapping the global model predictions to standard air quality classifications can have more impact on downstream tasks like exposure measurement and public health analytics. For example, rather than considering raw PM2.5 predictions from our meta-regression model, one can classify them using the Air Quality Index look up table\footnote{\url{https://www.epa.gov/wildfire-smoke-course/wildfire-smoke-and-your-patients-health-air-quality-index}}, to indicate the severity of exposure over a fixed period of time. To evaluate this approach, the accuracy of this classification can be assessed by comparing with classifications of actual PM2.5 values using F1 Scores.

\paragraph{Auxiliary data sources: } \label{sub:aux}
A major motivation of this project was to understand how lockdown policies in different countries have affected the exposure of children to air pollutants like PM2.5 and NO$_2$. In response to this objective, future versions of our model could benefit from the incorporation of lockdown-related input features.  We originally intended to use features from the ACAPS COVID-19 Government Measures Dataset\footnote{\label{foot:ACAPS}\url{https://www.acaps.org/covid-19-government-measures-dataset}} along with feature rich Oxford COVID-19 Government Response Tracker (OxCGRT)\footnote{\label{foot:OxCGRT}\url{https://github.com/OxCGRT/covid-policy-tracker}}, but due to data heterogeneity and the absence of location-specific measures across different locations, we removed them from the final model. Some examples of such features are related to lockdown dates and severity, social distancing, movement restrictions, relevant social, economic or public health measures, and the COVID-19 Government Response Stringency Index\footnote{\label{foot:Stringency}\url{https://ourworldindata.org/grapher/covid-stringency-index}}.

To account for local variability within the global model predictions, we intend to incorporate land cover \cite{landcover1,landcover2,landcover3}, and meteorological variables such as temperature, wind speed, wind direction, and cloud cover \cite{meteoro1,meteoro2}. We can capture these information using publicly available data sources such as the Copernicus yearly land cover classification \cite{landdata} and observational meteorological data collected by the United States Environmental Protection Agency (US EPA) \footnote{\url{https://www.epa.gov/scram/meteorological-data-and-processors}}.

The effect of population on children's respiratory health is well-documented. For example, in cities with populations more than 10 million, children are exposed to air pollution levels that are $2-8$ times higher than WHO-acceptable limits \cite{popeffect}. Through UNICEF's ties with local governments, we have obtained school locations in Niger, Sierra Leone, and Colombia and global child population data. We intend to augment our model with input features from these data \cite{Rees} to prioritize locations with higher child population and high predicted PM2.5 levels for post-COVID-19 monitoring. Another potential contributor to children's exposure to high levels of air pollution is {\it intra-day} variability. Alternative methods of feature aggregation may be considered such as aggregating PM2.5 values using regional travel behaviour patterns to account for daytime (such as rush-hour periods) and nighttime discrepancy in pollutant levels, as well as the variation at school locations and residential areas.

\section{Conclusion}
Due to the lack of ground-based sensors and the heterogeneity of sensor distribution around the world, there is a need to develop a global model of air pollution to understand the effects of the COVID-19 lockdowns. Through this work, we developed an initial model along with the expert comments from a multi-disciplinary panel, and charted a roadmap for progress towards the broader goal of developing locally-tuned models through iterative feedback from relevant stakeholders. By presenting our motivation and findings to the CHI community, we wish to draw their attention to the issue of equitable global air quality monitoring in the post-COVID-19 world, and highlight the role of collaborative research teams using multi-disciplinary methods to develop this research further.
\begin{acks}
This work is an outcome of the research and development undertaken through the Solve For Good platform with UNICEF as the primary stakeholder. The authors wish to thank UNICEF Office of Innovation for the computational resources and funding support to present the work. We thank qAira \& PlumeLabs for the regional air quality data in Peru. Additionally, one of the authors (PP) is supported by Visvesvaraya PhD Scheme of Ministry of Electronics \& Information Technology, Government of India, being implemented by Digital India Corporation and Fulbright Fellowship.
\end{acks}

  \bibliographystyle{ACM-Reference-Format}
  \balance
  \bibliography{0_template_double_sigchi}

\end{document}